\documentclass[11pt,a4paper]{article}
\usepackage{jcappub}
\usepackage{ifthen}
\usepackage{epsfig}
\usepackage{booktabs}

\newcommand{\dd}{\textrm{d}}

\newcommand{\beqra}{\begin{eqnarray}}
\newcommand{\eeqra}{\end{eqnarray}}
\newcommand{\beq}{\begin{equation}}
\newcommand{\eeq}{\end{equation}}
\newcommand{\n}{\ensuremath{\mathbf{n}}}

\title{Cosmological parameter estimation: impact of CMB aberration}
\author[a]{Riccardo Catena}
\author[b]{and Alessio Notari}
\affiliation[a]{Institut f\"ur Theoretische Physik, Friedrich-Hund-Platz 1, 37077 G\"ottingen, Germany} 
\affiliation[b]{Departament de F\'isica Fondamental i Institut de Ci\'encies del Cosmos,
Universitat de Barcelona, Mart\'i i Franqu\'es 1, 08028 Barcelona, Spain
}

\emailAdd{riccardo.catena@theorie.physik.uni-goettingen.de}
\emailAdd{notari@ffn.ub.es}
\abstract{The peculiar motion of an observer with respect to the CMB rest frame induces an apparent deflection of the observed CMB photons, {\it i.e.} aberration, and a shift in their frequency, {\it i.e.} Doppler effect. Both effects distort the temperature multipoles $a_{\ell m}$'s via a mixing matrix at {\it any} $\ell$. The common lore when performing a CMB based cosmological parameter estimation is to consider that Doppler affects only the $l=1$ multipole, and neglect any other  corrections.  In this paper we check the validity of this assumption in parameter estimation for a Planck-like angular resolution, both for a full-sky ideal experiment and also when sky cuts are included to model CMB foreground contaminations with a sky fraction similar to the Planck satellite. Assuming a simple fiducial cosmological model with five parameters, we simulated CMB temperature maps of the sky and added aberration and Doppler effects to the maps. We then analyzed with a MCMC in a Bayesian framework the maps with and without aberration and Doppler effects in order to assess the ability of reconstructing the parameters of the fiducial model. Although the correction on the power spectrum ${C_\ell}$ is larger than the cosmic variance at $\ell>1000$ and potentially important, we find that the bias on the parameters is negligible for Planck.}

\begin{document}
\maketitle

\section{Introduction}

The Cosmic Microwave Background is used as a fundamental tool to test Cosmological Models and to quantitatively extract the parameters of such models. This is usually done by extracting the Temperature  and the Polarization power spectra from the maps and by fitting them with a $\Lambda$CDM background model with an almost scale-invariant gaussian spectrum of density perturbations. However, if we observe the CMB with a large velocity $\beta\equiv v/c$ relative to such background, the image undergoes distortions due to the Doppler effect and due to the aberration effect and it is natural to wonder whether there is any sizable bias on such spectra and so also on cosmological parameter estimation due to the fact that we are not at rest with the CMB. 

It is always commonly assumed that the only sizable effect of an observer's velocity on the CMB map is the generation of a large dipole and any other possible effect is commonly ignored in all analyses. In fact, if we  consider a perfectly homogeneous map,  there is a Doppler effect  which induces a dipole of order $\beta$, while higher $n$-th multipoles are of order $\beta^n$. Since we observe a dipole of order $10^{-3}$, while other multipoles are  ${\cal O} (10^{-5})$, the standard lore is to interpret this large dipole as due to our velocity (which is then inferred to be $\beta=(1.23 \pm 0.003)\times 10^{-3}$~\cite{Lineweaver:1996xa}), and otherwise to ignore completely any other effect of our velocity.

However this would be true only on a perfectly homogenous sky. In reality we have a nontrivial map with nonzero coefficients $a_{\ell m}$, which are measured to be of order $10^{-5}$ for any $\ell>1$. Applying a Doppler effect induces a mixing between multipoles~\cite{Challinor:2002zh,Kosowsky:2010jm,Amendola:2010ty}. Precisely there is a mixing at order $\beta$ between neighboring multipoles $\ell$ and $\ell\pm1$ and a mixing at order $\beta^n$ among $\ell$ and $\ell\pm n$. Moreover there is another  important distortion, that is aberration, which changes the direction of observation of incoming photons, similarly to what lensing does. Aberration~\cite{Challinor:2002zh,Kosowsky:2010jm,Amendola:2010ty} can be computed for small $\beta$ and induces an additional coupling between neighboring multipoles of order $(\beta \ell)^n$. Such a distortion is therefore even more important because it becomes large at high $\ell$. Note that this is in fact the most important secondary effect on the CMB and it is not clear {\it a priori} if it induces a negligible bias on the power spectra and on the cosmological parameters: an ${\cal O}(\beta)$ or ${\cal O}(\beta \ell)$ effect on all multipoles could a priori lead to a large effect, when summed over all $\ell$ and $m$'s. Moreover for aberration this effect grows as $\beta \ell$ and becomes dominant at large $\ell$: using the value $\beta=1.23 \times 10^{-3}$ the distortion is order 1 on a single coefficient already at $\ell\gtrsim 800$! What happens in fact is that the aberration is deflecting the arrival direction of the photons by an angle of order $\beta$ and so if we are looking at angles smaller than $\beta$ there is an order 1 effect on the CMB.

However, a negative result came from~\cite{Challinor:2002zh} where it was found that, defining as usual $C_\ell\equiv 1/(2 \ell +1) \sum_m a^*_{\ell m} a_{\ell m}$, the bias on the ensemble average $\langle C_{\ell} \rangle$ undergoes some cancellations leading to a very small deviation which does not grow with $\ell$, precisely $\delta \langle C_{\ell} \rangle/ \langle C_{\ell} \rangle={\cal O} (\beta^2)$ at leading order in $\beta$.

More recently instead it has been realized that such conclusion is not the final word for several reasons. First of all such calculations are based on a perturbative approach in $\beta$ and so, as we mentioned before, they apply only up to $\ell \lesssim 800$, while we do not know if a clear bias could show up at higher $\ell$'s~\cite{Amendola:2010ty,Notari:2011sb}. Second,  such an estimate is based on a full-sky definition for $C_\ell$ : in fact, in full-sky, because of the sum over $m$ there are cancellations between the corrections on the individual $a_{\ell m}$'s.  However we never observe a full-sky in an experiment and it can be easily seen that as long as one part of the sky is removed such cancelllations do not hold and we can get again effects of order $\beta \ell$, which are larger, on the pseudo-$C_{\ell}$'s which have to be used in a partial sky estimate. This was noticed in~\cite{Pereira:2010dn}, where the bias on the pseudo-$C_{\ell}'s$ has been analyzed for small $\beta \ell$ and for very low $\ell$, less than about 20. This estimate becomes very expensive computationally at higher $\ell$ and anyway not feasible within pertubation theory at large $\ell \gtrsim 1/\beta$ . For all these reasons the question whether the effect of velocity is negligible on the power spectrum is still lacking an answer.

The scope of the present paper is to resolve this issue by analyzing the bias on cosmological parameter estimation due to a nonzero $\beta$ at {\it any} $\ell$ and for both full-sky and cut-sky. As we have stressed this could be done perturbatively analytically only for low-$\ell$, while another possibility, valid at any $\ell$, would be to write down the exact mixing matrix between {\it any} multipole without pertubative expansions. This would amount to numerically compute integrals of highly oscillating Legendre polynomials, which becomes very slow at high-$\ell$, and becomes extremely slow for a cut-sky, since it involves also a window function. This approach has been addressed by recursive formulae in~\cite{Chluba:2011zh} and by approximate fitting functions in~\cite{Notari:2011sb}.

The way we proceed here is a more direct and straightforward one: we simulate maps of the CMB sky and we directly apply on the maps the boost trasformation before extracting the $a_{\ell m}$'s, therefore bypassing the need of computing the  mixing coefficients. Then we extract the $a_{\ell m}$'s and the $C_{\ell}$'s and finally we run a Markov Chain Monte Carlo (MCMC) to estimate cosmological parameters and evaluate the bias between the moving  observer and the rest frame observer. This approach also allows us to easily implement sky-cuts and to check whether the effect is larger on a realistic cut-sky experiment. We apply all this procedure to Temperature maps only and both for WMAP-like resolution and for Planck-like resolution.

The paper is organized as follows: in section \ref{ab} we review the impact of aberration and Doppler on CMB observables. Section \ref{data} is instead devoted to a detailed description of the procedure followed to simulate CMB temperature maps properly including Doppler and aberration effects. These maps are then analyzed in section \ref{mcmc}. In section \ref{fitting} we check that using fitting functions from~\cite{Notari:2011sb} we can reconstruct the aberrated $C_{\ell}$'s analytically.
Finally section \ref{conclusion} summarizes our results.

\section{CMB aberration and Doppler}
\label{ab}

For an observer in motion with respect to a light source, the aberration phenomenon consists in the apparent deflection of the observed light bundles due to the relative motion between the observer and the source. In the case of the cosmic microwave background (CMB), as a consequence of aberration, an observer moving with respect to the CMB rest frame assigns to a photon emitted along the direction $\hat{\n} $ at the last scattering surface the arrival direction ${\hat{\n}'}$, where $\hat{\n}$ and $\hat{\n}'$ are unit vectors. The aberration angle ${\bf \alpha} \equiv \hat{\n}' - \hat{\n}$ can be calculated from the velocity transformation relating the CMB and the observer reference frames: assuming that the observer moves along the direction of the z-axis (identified by the unit vector $\hat{\mathbf{z}}$) with speed $\beta$ compared to the CMB reference frame, one finds
\begin{equation}
{\bf \alpha} \cdot \hat{\mathbf{z}} \;= \frac{\beta \sin^2\theta }{1+\beta  \cos\theta} \,
\label{aberration}
\end{equation}
where $\cos\theta = \hat{\bf n}\cdot\hat{\bf z}$. We refer to the relative motion between the observer and the CMB rest frame as the peculiar motion of the observer. Besides an aberration of the CMB radiation, a peculiar motion of the observer induces also a change in the frequency of the observed photons. According to Special Relativity the frequency $\nu'$ in the moving frame is related to the frequency $\nu$ measured at rest by the usual Doppler effect:
\begin{equation}
\nu'=\nu\gamma(1+\beta\hat{\mathbf{n}}\cdot\hat{\mathbf{z}})\,
\end{equation}
where $\gamma = 1/\sqrt{1-\beta^2}$. Doppler and aberration effects unavoidably distort the CMB temperature and polarization maps currently measured by experiments such as for instance the WMAP or Planck satellites. This can be seen as follows: CMB experiments directly measure the brightness of the observed radiation. The brightness $I(\nu)$ is then conventionally translated into an equivalent value for the thermodynamic temperature $T$ through the black body law
\begin{equation}
  T = \frac{I(\nu)}{\nu^2}\frac{e^{\nu/T}-1}{\nu/T} \,.
\label{T}
\end{equation}
By solving Eq.~(\ref{T}) one finds
\begin{equation}
T = \frac{\nu}{\log\left(1+\frac{\nu^3}{I(\nu)}\right)} \,
\label{T2}
\end{equation}
which is therefore the temperature indirectly measured by CMB experiments. Analogously, in the case of the CMB polarization one translates the quantities actually measured into the value of an effective temperature obtained replacing in Eq.~(\ref{T2}) the brightness $I(\nu)$ with the relevant Stokes parameters. Since the ratio $\nu^3/I(\nu)$ is invariant with respect to frame transformations \cite{Challinor:2002zh}, one can conclude from Eq.~(\ref{T2}) that under a Lorentz boost the temperature has to transform as follows
\begin{equation}
T^{\prime}(\hat{\mathbf{n}}') =  \gamma(1+\beta\hat{\mathbf{n}}\cdot\hat{\mathbf{z}}) T(\hat{\mathbf{n}}) \,
\label{boost}
\end{equation}
where $T^{\prime}$ is the temperature in the moving frame and $T$ is the temperature measured by an observer at rest (with respect to the CMB). Eq.~(\ref{boost}) together with Eq.~(\ref{aberration}) thus describes how a temperature map is distorted consequently to aberration and Doppler effects: the temperature associated with the direction $\hat{\n}' $ by an observer in motion with peculiar velocity $\beta{\bf \hat{z}}$ is equal to the temperature that an observer at rest would assign to the direction $\hat{\n}$, multiplied by a corrective Doppler factor which depends on $\beta$. Expanding in spherical harmonics both $T^{'}(\hat{\n}') = \sum_{lm} a'_{lm} Y_{lm}(\hat{\n}')$ and $T(\hat{\n})= \sum_{lm} a_{lm} Y_{lm}(\hat{\n})$, Eq.~(\ref{boost}) can be rewritten as follows\footnote{Analogous expressions for the CMB Stokes parameters involves spin-weighted spherical harmonics.}  
\begin{equation}
  a'_{\ell^{\prime}m^{\prime}} = \sum_{\ell} \int d\hat{\n}\,a_{\ell m'}
\left[\gamma(1+\beta\hat{\n}\cdot{\hat{\bf z}})\right]^{-2}Y_{\ell^{\prime}m^{\prime}}^{*}(\hat{\n}^{\prime})Y_{\ell
m^{\prime}}(\hat{\n})\,.\label{boost2}
\end{equation}
This expression can be then used to compute the desired correlation functions and compare these results with observations, properly accounting for aberration and Doppler effects. Eq.~(\ref{boost2}) is exact, though it relies on an integral computationally very expensive for high multipoles.

In the following subsections we will review two approximations recently discussed in the literature to estimate the coefficients $a'_{lm}$ of Eq.~(\ref{boost2}). In sections \ref{data} and \ref{mcmc}, instead, we will show how starting from Eq.~(\ref{boost}) one can actually include aberration and Doppler effects {\it exactly} in a CMB based cosmological parameter estimation, by transforming directly simulated maps of the sky in the boosted frame.

\subsection{Perturbative expansions}
Since our peculiar velocity is small in natural units, {\it i.e.} $\beta \simeq 1.23\times 10^{-3}$,  Eq.~(\ref{boost2}) can be evaluated expanding the integrand around $\beta$ equal to zero.  The leading corrections in $\beta$ necessary to evaluate the $C_{\ell}$'s are given by
\begin{equation}
  a'_{\ell\, m}\simeq c_{\ell m}^{-}a_{\ell-1\, m}+c_{\ell m}^{+}a_{\ell+1\, m}+  a_{\ell\, m} (1+d_{\ell m})\,
  \label{order1}
\end{equation}
where
\begin{eqnarray}
  c_{\ell m}^{+} & = & \beta(\ell+1) \, G_{\ell+1, m} \,,\nonumber \\
  c_{\ell m}^{-} & = & -\beta\ell \, G_{\ell, m} \, ,\nonumber \\
  d_{\ell m} & = & \frac{\beta^2}{2} \left[ (\ell+1) (\ell+2) \, G_{\ell+1, m}^2 + \ell (\ell-1) \, G_{\ell, m}^2  - \ell (\ell+1)+ m^2-1\right]
\label{Ccoef}
\end{eqnarray}
and
$G_{\ell, m}\equiv \sqrt{\frac{\ell^{2}-m^{2}}{4\ell^{2}-1}}$.
Similar expressions can be derived for the analogous coefficients corresponding to the Stokes parameters describing the CMB polarization. We refer the reader to Refs. \cite{Challinor:2002zh,Kosowsky:2010jm,Amendola:2010ty,Notari:2011sb} for an explicit derivation of such expressions; note that the correct expressions for Temperature fluctuations for   the $c_{\ell m}^{\pm}$ have been derived first in~\cite{Kosowsky:2010jm} and for $d_{\ell m}$  in \cite{Notari:2011sb} and they slightly differ from \cite{Challinor:2002zh}, which would be valid for Intensity instead of Temperature. We note here that the corrections increase as powers of $\beta\ell$, and so become large for high $\ell$ and so for $\ell>1/\beta$ the perturbative approach breaks down. This happens because the aberration effect is a distortion on the angle of order $1/\beta$, and this is not a small correction if we are looking at angular scales smaller than $1/\beta$. Starting from Eq.~(\ref{order1}) one can easily compute the desired correlation functions. In the case of the angular power spectrum $C^{\prime}_{\ell} \equiv  \langle a^{\prime}_{\ell m}  a^{\prime *}_{\ell m}  \rangle$, for instance, one obtains 
\begin{eqnarray}
C^{\prime}_{\ell} &=& \sum_{\ell'} C_{\ell'}\Bigg\{ \delta_{\ell \ell'} \left(1- \frac{1}{3}\beta^2(\ell^2+\ell+1)\right)
+ \delta_{\ell(\ell'+1)}\beta^2 \frac{\ell^3}{3(2\ell+1)} \nonumber\\
&+& \delta_{\ell(\ell'-1)} \beta^2\frac{(\ell+1)^3}{3(2\ell+1)} \Bigg\}\,.
\label{Cl}
\end{eqnarray}

Thus, for $l<1/\beta$ the leading corrections to the primordial angular power spectrum $C_{\ell} \equiv  \langle a_{\ell m}  a^{*}_{\ell m}  \rangle$ are small, {\it i.e.} $\mathcal{O}(\beta^2 l^2)$, and moreover they are subject to a further suppression since they would give zero in the limit of flat angular power spectrum when $C_\ell \sim C_{\ell+1}$ \footnote{Note that such correction goes to zero only if computed with the correct coefficients  Eq.~(\ref{order1}), derived in \cite{Notari:2011sb} instead of going to $4 \beta^2$, as previously obtained by \cite{Challinor:2002zh}. }.  This is the reason why in the literature the aberration has been not considered so far in the cosmological parameter probabilistic inference: according to Eq.~(\ref{Cl}) the aberration only mildly affects the observed $C^{\prime}_l$ and can be therefore safely neglected studying with standard MCMC methods the Bayesian credible regions and frequentist confidence levels of the cosmological parameters. However as already stressed in the Introduction, this is not completely safe for two reasons.  First of all, when going at $\ell \gtrsim 1/\beta$ the corrections due to aberration and Doppler are not under perturbative control and could turn out to be large also for the $C_{\ell}$'s. Second,  Eq.~(\ref{Cl}) and (\ref{Ccoef}) are correct only in the full sky approximation, which does not account for any foreground contamination of the observed maps. In a more realistic computation, in fact, one has to consider that indeed only a fraction of the sky is experimentally accessible. This can be done by multiplying the original full sky map by an opportune window function to mask the background dominated portions of the sky and this might to a larger bias in the $C_{\ell}$'s: in fact the small result obtained in~(\ref{Cl}) can be understood as a cancellation of a larger effect when summing over the sphere, but such cancellation does not hold when we look only at a portion of the sky.
More precisely: (1) an effect of $\mathcal{O}(\beta l)$ shows up~\cite{Pereira:2010dn} for asymmetric maps (2) even for symmetric maps the above mentioned suppression which acts  when $C_\ell \sim C_{\ell+1}$  becomes less effective, because the sum involves a larger number of $C_{\ell}$'s and not only the neighbours. In the present paper we analyze only symmetric maps and we leave to further work the analysis with asymmetric maps.

The procedure of applying a window function leads to a fit of the data based on the concept of pseudo-$C_l$, which we review  in the next section. We conclude this subsection mentioning that, because of aberration and Doppler effects, not only the primordial $C_l$ are distorted, but also the off-diagonal terms of the covariance matrix ({\it i.e} $\langle a'_{\ell_1 m}a'^*_{\ell_2 m}\rangle$ with $\ell_1\neq\ell_2$). In a series of recent work it has been shown that this property of the off-diagonal terms can be used to effectively measure the magnitude and direction of our peculiar motion. In Ref.~\cite{Amendola:2010ty}, for instance, it has been found perturbatively that an experiment such as Planck will be able to measure $\beta$ with an accuracy of about 30\%, and the direction of this motion with an error of about 20 degree, while Ref.~\cite{Notari:2011sb} shows non-perturbatively that such accuracy would be of about 20\% for Planck and down to 10\% or even 5\% for future planned experiments. Perturbative corrections  to the covariance matrix of the $a'_{lm}$ were also calculated beyond the first order approximation. A computation of  $\langle a'_{\ell_1 m}a'^*_{\ell_2 m}\rangle$ at second order in $\beta$ can be found in  Refs.~\cite{Challinor:2002zh, Notari:2011sb}, while higher order corrections have be calculated in some cases by \cite{Notari:2011sb} and by using recurrence formulas in Ref.~\cite{Chluba:2011zh}.

\subsection{Fitting formulas}
Eq.~(\ref{boost2}) can be rewritten in matrix form as follows
\begin{equation}
  a'_{\ell m}\; =\; \sum_{\ell'} K_{\ell' \, \ell\, m} \, a_{\ell' m}\,
  \label{matrixalm}
\end{equation}
where the kernel $K_{\ell' \, \ell\, m}$ is equal to
\begin{equation}     
 K_{\ell'\, \ell\, m} = \int_{-1}^{1}  \frac{\dd x}{\gamma\, (1-\beta x)} \,\tilde{P}_{\ell'}^m(x) \, \tilde{P}_{\ell}^m \! \left(\frac{x - \beta}{1 - \beta  x}\right) \,
 \label{abkernel}
\end{equation}
and 
\begin{equation}
  \tilde{P}_{\ell}^m(x) \;\equiv\; \sqrt{\frac{2\ell+1}{2} \frac{ (\ell-m)!}{(\ell+m)!}} \,P_{\ell}^m(x) \,.
\end{equation}
 The $P_{\ell'}^m(x)$ are the associated Legendre polynomials. The $ K_{\ell'\,\ell\,m}$  (the so-called aberration kernel), which relates the aberrated coefficients $a'_{\ell m}$ to the primordial ones $ a_{\ell m } $, are of course given at leading order  in $\beta \ell \ll 1$  by expressions such as eq.~(\ref{Ccoef}), but it is more problematic to compute them at large $\ell$ beyond a perturbative approach. Such problem has been recently addressed in \cite{Notari:2011sb} proposing an approximate solution to the oscillating integrals of Eq.~(\ref{boost2}). In this paper simple fitting formulas were constructed which, approximating the oscillatory behavior of the relevant integrals by appropriate Bessel functions, reproduce the exact result (verified numerically in a few representative cases) for the elements  of the aberration kernel with an accuracy of about 0.1 per cent. For example, for the temperature aberration kernel an accurate fitting formula is
\begin{eqnarray}     
  K_{\ell-1\, \ell\, m}   &&\;\simeq\; J_1\!\Big(\! -2\, c^{-}_{\ell\, m} \Big) \\
  K_{\ell+1\, \ell\, m}   &&\;\simeq\; J_1\!\Big( \,2\, c^{+}_{\ell\, m} \Big)
\end{eqnarray}
where $J_1$ is a Bessel function of the first kind. Similar expressions also exist for the polarization aberration kernels. We refer the reader to \cite{Notari:2011sb} for a complete list of fitting formulas and a detailed discussion of the derivation and regime of validity. In the next section, we will go beyond the approximation schemes discussed in these subsections showing how to include exactly aberration and Doppler effects in a MCMC scan of the cosmological parameter space. 

\section{Simulated maps including aberration and Doppler}
\label{data}
The ultimate goal of this analysis is to determine the impact of aberration and Doppler on a CMB based cosmological parameter estimation. This requires a method to simulate CMB maps including aberration and Doppler, in particular when windows functions are considered in the analysis to model the CMB maps foreground contamination. Mock data accounting for these phenomena can be generated and analyzed by using a straightforward generalization of the pseudo-$C_\ell$ approach. In the following, after introducing the basic formulas to compute the pseudo-$C_\ell$ including aberration and Doppler, we will discuss their implementation in a modified version of the HEALPix\footnote{http://healpix.jpl.nasa.gov} code \cite{Gorski:2004by}. This numerical tool will allow us to generate mock data measured by an observer moving with respect to the CMB rest frame. The next sections will be then devoted to a MCMC analysis of these simulated data.  

\subsection{The boosted pseudo-$C_{\ell}$}
To simplify the expression of the aberration kernel (see Eq.~(\ref{abkernel})), we assumed so far that the direction of the observer peculiar motion coincides with ${\bf \hat{z}}$, namely the direction of the $z$-axis. However, when dealing with window functions, it is convenient to work using Galactic coordinates and assume therefore that the $z$-axis identifies a direction perpendicular to the Galactic plane. We know however that our peculiar motion does not occur perpendicularly to the Galactic plane, but it is instead associated with the direction of the CMB dipole, measured along the direction identified by the longitude $l\simeq 249$ and the latitude $b\simeq 48$. This implies that to express the coefficients computed in Eq.~(\ref{boost2}) consistently with our observed peculiar motion, we need to rotate them with a Wigner matrix $\mathcal{D} _{m m'}^{\ell}(\phi,\theta,\gamma) $, where $\phi,\theta,\gamma$ are the Euler angles required to align the peculiar velocity, initially along ${\bf \hat{z}}$, with the CMB dipole. This Wigner rotation reads as follows
\beq
\hat{a}_{\ell m} = \sum_{-\ell \le m' \le \ell} \mathcal{D}_{m m'}^{\ell}(\phi,\theta,\gamma) \,a'_{\ell m'} 
\label{rot}
\eeq
where the coefficients $a'_{\ell m}$ were calculated in Eq.~(\ref{boost2}). The coefficients $\hat{a}_{\ell m}$ are now consistently expressed in the frame of an observer moving with velocity $\vec{\beta}$ in the direction of the CMB dipole. These equations are correct in a full sky harmonic decomposition. However, current CMB experiments observe only a portion of the whole CMB sky. From a theoretical point of view, this limitation can be modeled through opportune windows functions which set to zero the contributions to the computed quantities coming from regions of the sky background dominated ({\it e.g.} the Galactic plane). A window function $W({\bf \hat{n}})$ acts multiplicatively on the original full sky map. Therefore, if we denote by $\hat{T}({\bf \hat{n}})$ a full sky temperature map, its cut sky counterpart $\tilde{T}({\bf \hat{n}})$ reads as follows
\beqra
\tilde{T}({\bf \hat{n}}) &=& W({\bf \hat{n}}) \hat{T}({\bf \hat{n}}) \nonumber\\
&=& \sum_\ell \sum_{-\ell\le m\le\ell} w_{\ell m}Y_{\ell m}({\bf \hat{n}}) \hat{T}({\bf \hat{n}}) 
\eeqra
where in the second line we expanded the window function in spherical harmonics and denoted by $w_{\ell m}$ the coefficients of this decomposition. A further expansion of the full sky map $\hat{T}(\hat{\n})= \sum_{lm} \hat{a}_{\ell m} Y_{lm}(\hat{\n})$  and of the cut sky map $\tilde{T}(\hat{\n})= \sum_{\ell m} \tilde{a}_{\ell m} Y_{\ell m}(\hat{\n})$ finally leads to  
\beq
\tilde{a}_{\ell_{1}m_{1}} = \sum_{\ell_ {2}}\sum_{-\ell_{2}\le m_{2}\le\ell_{2}} \mathcal{F}_{\ell_{1}m_{1}\ell_{2}m_{2}} \,\hat{a}_{\ell_{2}m_{2}} 
\label{window}
\eeq
where the kernel $\mathcal{F}_{\ell_{1}m_{1}\ell_{2}m_{2}} $ relating the full sky coefficients $\hat{a}_{\ell_{2}m_{2}}$ to the cut sky coefficients $\tilde{a}_{\ell_{1}m_{1}}$ admits the following representation in terms of the Wigner 3-$j$ symbols
\beqra
 \mathcal{F}_{\ell_{1}m_{1}\ell_{2}m_{2}} &=& \sum_{\ell_ {3}}\sum_{-\ell_{3}\le m_{3}\le\ell_{3}} 
w_{\ell_{3} m_{3}} (-1)^{-m_{2}} \sqrt{\frac{(2\ell_{1}+1)(2\ell_{2}+1)(2\ell_{3}+1)}{4\pi}} \nonumber\\
\nonumber\\
&\times& 
 \left( \begin{array}{ccc}
 \ell_1 & \ell_2 & \ell_3 \\
 0 & 0 & 0 
 \end{array} \right)
 \left( \begin{array}{ccc}
 \ell_1 & \ell_2 & \ell_3 \\
 m_1 & -m_2 & m_3 
 \end{array} \right) \,.
\eeqra
In the present analysis the $\hat{a}_{\ell m}$ coefficients are the ones given in Eq.~(\ref{rot}). The coefficients $\tilde{a}_{\ell m}$ are then used to construct the pseudo-$C_{\ell}$, which are denoted by $\tilde{C}_{\ell}$ and defined as follows 
\beq
\tilde{C}_{\ell} \equiv \frac{1}{2\ell+1} \sum_{-\ell \le m \le \ell} |\tilde{a}_{\ell m}|^2 \,.
\label{pseudo}
\eeq
Combining now Eqs.~(\ref{window}) and (\ref{pseudo}), we can finally obtain the desired expression for the pseudo-$C_{\ell}$ which includes also aberration and Doppler. For the ensemble average of the $\tilde{C}_{\ell}$ one finds
\beq
\langle\tilde{C}_{\ell_{1}}\rangle =\frac{1}{2 \ell_1+1} \sum_{-\ell_{1}\le m_{1}\le\ell_{1}} \sum_{\ell_{2} m_{2}} \sum_{\ell_{3} m_{3}} 
\mathcal{F}_{\ell_{1}m_{1}\ell_{2}m_{2}} \mathcal{F}^{*}_{\ell_{1}m_{1}\ell_{3}m_{3}} 
\,\langle \hat{a}_{\ell_{2} m_{2}} \hat{a}^{*}_{\ell_{3} m_{3}} \rangle \,.
\label{pseudo2}
\eeq
An analytic expression for $\langle \tilde{C}_{\ell_{1}} \rangle$ at first order in $\beta$ has been obtained in Ref.~\cite{Pereira:2010dn}. In the limit $\beta=0$ and assuming statistical isotropy Eq.~(\ref{pseudo2}) reduces to 
\beq
\langle\tilde{C}_{\ell_{1}}\rangle = \sum_{\ell_{2}} \mathcal{M}_{\ell_{1}\ell_{2}} C_{\ell_{2}} 
\label{pseudo3}
\eeq
where the mode-coupling matrix $\mathcal{M}_{\ell_{1}\ell_{2}}$ reads as follows
\beq
\mathcal{M}_{\ell_{1}\ell_{2}} = \frac{2\ell_{2}+1}{4\pi}\sum_{\ell_{3}} (2\ell_{3}+1) \mathcal{W}_{\ell_{3}}
\left( \begin{array}{ccc}
\ell_1 & \ell_2 & \ell_3 \\
0 & 0 & 0 
\end{array} \right)^{2}
\eeq
and $\mathcal{W}_{\ell_{3}} = 1/(2\ell_{3}+1) \sum_{m_{3}} |w_{\ell_{3}m_{3}}|^2$ is the window function power spectrum. The pseudo-$C_{\ell}$ formalism is very convenient to construct estimators of the full sky $C_{\ell}$. The best estimate of the ensemble average $\langle\tilde{C}_{\ell}\rangle$ is given by $\tilde{C}_{\ell}^{\textrm{exp}}$, the value of $\tilde{C}_{\ell}$ actually observed or extracted from simulations. Then, by inverting the mode-coupling matrix $\mathcal{M}_{\ell_{1}\ell_{2}}$, one can construct from Eq.~(\ref{pseudo3}) an unbiased estimator, denoted by $C^{\dagger}_{\ell}$, of the full sky $C_\ell$ 
\beq
C^{\dagger}_{\ell_{1}}=\sum_{\ell_{2}} \mathcal{M}^{-1}_{\ell_{1}\ell_{2}} \tilde{C}_{\ell_{2}}^{\textrm{exp}} \,.
\label{estimator}
\eeq
Now, to extract an estimator of the full sky $C_{\ell}$ from Eq.~(\ref{pseudo2}) is instead extremely more complicated than in Eq.~(\ref{pseudo3}), since the boosted $\tilde{C}_{\ell}$ involve the full covariance matrix instead of only its diagonal part. 
We will come back to this important observation in section \ref{mcmc} discussing its role in the cosmological parameter estimation. 

In the usual case of ignoring $\beta$, the goodness of estimators like Eq.~(\ref{estimator}) has been tested in various works \cite{Wandelt:2000av}. In the next sections, when referring to $C^{\dagger}_{\ell}$, we will use for simplicity an estimator obtained replacing in Eq.~(\ref{estimator}) the mode-coupling matrix with the $f_{sky}$ parameter (defined as the portion of the sky covered divided by $4\pi$). This choice has been shown to be fairly robust in the context of weak lensing analyses \cite{Das:2007eu} and it is accurate enough for the present investigation, which does not aim at proposing a new estimator to account for aberration but rather at determining the error induced by neglecting the peculiar motion of the observer when performing analyses based on Eq.~(\ref{estimator}) (with $ \mathcal{M}^{-1}_{\ell_{1}\ell_{2}} \rightarrow f^{-1}_{sky}$). 
\subsection{Numerical implementation}
\label{exp}
The equations introduced in the previous sections involve oscillatory angular integrals (see Eq.~(\ref{boost2})) and rotations in harmonic space ({\it e.g.} Eqs.~(\ref{window}) and  (\ref{rot})). To handle these complications we will make use of a modified version of the HEALPix code~\cite{Gorski:2004by}, a numerical package to sample functions defined on the sky through a hierarchical, equal area, iso-latitude pixelisation of the sphere. The aim is to simulate CMB data including aberration and Doppler. This requires a fiducial cosmological model for which the angular power spectrum $C_{\ell}^{(f)}$ is assumed to be known. Our choice will be specified at end of this section. The data simulation is then articulated in five steps which for clarity and completeness we list here in the following together with the HEALPix routines used in the generation of the mock data.
\begin{itemize}
\item {\it synfast}: The synfast program generates full sky maps sampling the corresponding harmonic coefficients $a_{\ell m}$ from a Gaussian distribution with zero mean and variance equal to the $C_{\ell}^{(f)}$. We modified this code including the possibility of applying the transformation law given in Eq.~(\ref{boost}) directly on the pixels. In this way the aberration and Doppler effects are included directly in position space rather than in harmonic space. As a consequence, one can avoid the evaluation of the complicated oscillatory integral in Eq.~(\ref{boost2}). 
We set the map resolution parameter $N_{\textrm{side}}$ to the value 2048.  
\item {\it anafast}: We analyze the temperature map obtained in the previous step with the standard HEALPix version of the anafast code, expanding it in spherical harmonics and determining the corresponding multipole coefficients. These are the {\it exact} ({\it i.e.} non perturbative) $a'_{\ell m}$ coefficients of Eq.~(\ref{boost2}). 
\item {\it alteralm}: We also modified the alteralm program in order to rotate these $a'_{\ell m}$ in a new reference frame where the peculiar velocity inducing the aberration has the direction of the CMB dipole (this transformation is not implemented in the the standard version of alteralm). This corresponds to evaluate Eq.~(\ref{rot}).  
\item {\it synfast}: Then we performed a second call to the synfast program to generate the temperature map $\hat{T}({\bf \hat{n}})$ associated with the rotated $\hat{a}_{\ell m}$ coefficients.
\item {\it anafast:} A final call to the anafast program allows to introduce the desired window function and obtain the observed $\tilde{C}_{\ell}^{\textrm{exp}}(\beta)$, which are in general a function of the peculiar velocity $\beta$.
\end{itemize}
This procedure provides the CMB datasets analyzed in the next section. The fiducial model used in this paper is similar to the one studied in Ref.~\cite{Hamimeche:2008ai}. It corresponds to a toy cosmological model with five free parameters, namely the baryon density in units of the critical density times the square of the present value of the Hubble rate $\Omega_{b} h^2$, the analogous quantity for the dark matter component $\Omega_{dm} h^2$, the spectral index $n_s$, the amplitude of the primordial power spectrum $A_{s}$ and, finally, 100 times the ratio of the sound horizon to the angular diameter distance at recombination $\theta$. The fiducial model is characterized by the following parameter values: $\Omega_{b} h^2=0.022$, $\Omega_{b} h^2=0.12$, $n_s=1$, $A_s=2.3\times10^{-9}$ and $\theta$ has been set to a value corresponding to $h=0.7$. The other cosmological parameters have been fixed as in the file test$\_$params.ini which can be downloaded at the web page \cite{lewis}.  This simplified choice of the cosmological parameters allowed us to study the impact of aberration within a model for which the Likelihood approximation implemented in the next section has been carefully tested in Ref.~\cite{Hamimeche:2008ai}. 
\begin{figure}[t]
\begin{minipage}{50mm}
\includegraphics[width=70mm,height=50mm]{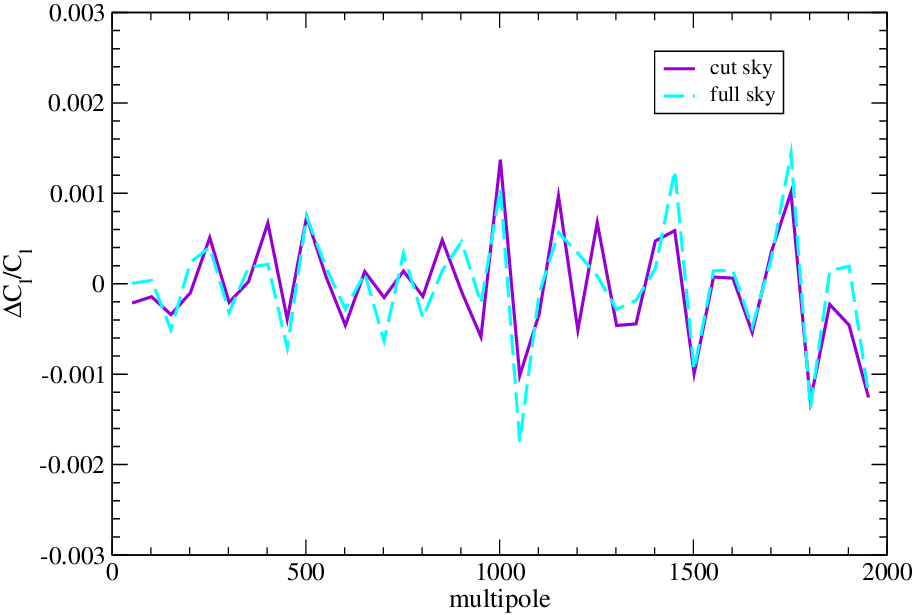}
\end{minipage}
\hspace{30mm}
\begin{minipage}{50mm}
\includegraphics[width=70mm,height=50mm]{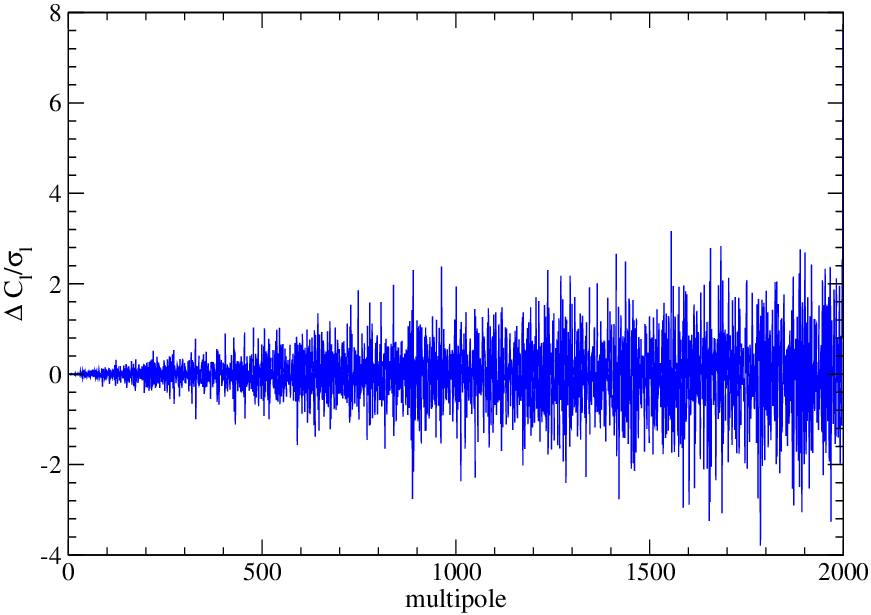}
\end{minipage}
\caption{Left panel: $\Delta C_{\ell}/C_{\ell}$ as a function of the multipole $\ell$ for a simulated temperature map with $f_{sky}=0.826$ and for the same simulation but in a full sky configuration. To emphasize the impact of CMB Doppler and aberration, $\Delta C_{\ell}/C_{\ell}$ has been binned with a bin size equal to 50. Right panel: $\Delta C_{\ell}/\sigma_{\ell}$ as a function of the multipole $\ell$ for a simulated temperature map with $f_{sky}=0.826$. On a single $\ell$ aberration and Doppler effects alter the pseudo-$C_\ell$ up to a few times the cosmic variance $\sigma_\ell$.}
\label{sim}
\end{figure}

\subsection{Datasets}
\label{mock}
Starting from 20 different HEALPix random seeds we simulated 20 pairs of temperature maps with Planck resolution ($l_{max}=2000$) and a sky cut of 20 degrees around the Galactic plane: within a pair of temperature maps, one map was generated in the CMB rest frame, while the second in a frame moving with $\beta = 1.23 \times 10^{-3}$, including Doppler and aberration following the procedure described in the previous subsection. Then for all these maps we calculated the angular power spectrum, {\it i.e.} $\tilde{C}_{\ell}^{\textrm{exp}}(\beta)$ in the moving frame and $\tilde{C}_{\ell}^{\textrm{exp}}(0)$ in the CMB rest frame, obtaining the datasets analyzed with the CosmoMC program in the next sections. For one pair of temperature maps, associated with a simulation labelled by $\mathcal{S}$, we also evaluated these angular power spectra in a full sky configuration, therefore without applying the mentioned azimuthal cut. 

In the left panel of Fig.\ref{sim} we show the relative difference $\Delta C_{\ell}/C_{\ell}\equiv1-\tilde{C}_{\ell}^{\textrm{exp}}(\beta)/\tilde{C}_{\ell}^{\textrm{exp}}(0)$ as a function of the multipole $\ell$ for the simulation $\mathcal{S}$, in a full sky configuration (cyan) and in a cut sky configuration (violet). For clarity $\Delta C_{\ell}/C_{\ell}$ has been binned with a bin size equal to 50. From this figure we can see that both curves oscillates between approximately $-1\times10^{-3}$ and $1\times10^{-3}$. In the right panel of Fig.\ref{sim} we finally compare $\Delta C_{\ell}\equiv \tilde{C}_{\ell}^{\textrm{exp}}(0)-\tilde{C}_{\ell}^{\textrm{exp}}(\beta)$ with the cosmic variance, which is given by $\sigma_{\ell}=\sqrt{2/(2l+1)f_{sky}}\tilde{C}_{\ell}^{\textrm{exp}}(0)$, for a cut sky configuration and without binning $\Delta C_{\ell}/\sigma_{\ell}$. In this figure one can see that for a given $\ell$ the effect is large, being even a few times the cosmic variance. 

In the following sections we will present the results of $2\times 20$ independent CosmoMC runs associated with all the maps at disposal. This will allow us to quantify the impact of aberration and Doppler on a CMB based parameter estimation ``averaging'' over the number of HEALPix seeds used in this work.  

\section{MCMC analysis}
\label{mcmc}
Let us assume that a CMB experiment has measured a certain temperature map, say $T^{\textrm{exp}}({\bf \hat{n}})$, with associated angular power spectrum $C^{\textrm{exp}}_{\ell}$. If we wanted to use this map to infer the underlying cosmological model, one possibility would be to use directly Eq.~(\ref{estimator}) to derive from $C^{\textrm{exp}}_{\ell}$ the corresponding estimator of the full sky angular power spectrum, $C^{\dagger}_{\ell}$, and implement this quantity in a Likelihood based analysis. This approach would be correct if the experiment were comoving with the CMB, but it would instead introduce a ``bias'' in the final parameter reconstruction if the observer were moving with a peculiar velocity different from zero, as one can deduce by comparing Eqs.~({\ref{pseudo2}}) and (\ref{pseudo3}). The common lore in this case is to {\it assume} that such a bias is negligible. In the following we will reconsider this assumption performing 2$\times$20 independent cosmological parameter estimations, one for every dataset considered in section \ref{mock}. We will also perform additional 2$\times$1 parameter estimations associated with the full sky power spectra related to the simulation $\mathcal{S}$. In these analyses we will {\it always} use the estimator $C^{\dagger}_{\ell}$, defined in Eq.~(\ref{estimator}) (with $ \mathcal{M}^{-1}_{\ell_{1}\ell_{2}} \rightarrow f^{-1}_{sky}$), as an experimental input for our Likelihood based reconstruction procedure (described in detail in the next sections). In this way, depending from which dataset we are considering, the resulting parameter reconstruction will be biased or not (in the sense explained above). Comparing then biased and unbiased reconstructions we can estimate the impact of a peculiar motion on the cosmological parameter estimation. Summarizing, there are two possible cases: 
\begin{itemize}
\item The dataset has been generated including aberration and Doppler. In this case an appropriate estimator of the full sky $C_{\ell}$ could be found solving Eq.~(\ref{pseudo3}) for the covariance matrix of the primordial $a_{\ell m}$. Using instead Eq.~(\ref{estimator}) in the Likelihood introduces an error in the analysis which consists in interpreting the data with a wrong theoretical assumption, namely assigning a null peculiar velocity to the observer. These considerations apply to the datasets $\tilde{C}_{\ell}^{\textrm{exp}}(\beta)$.
\item The dataset has been generated without introducing aberration and Doppler according to Eq.~(\ref{boost}). In this case an analysis based on Eq.~(\ref{estimator}) is perfectly correct. This applies to the datasets $\tilde{C}_{\ell}^{\textrm{exp}}(0)$.
\end{itemize}
Comparing therefore the results obtained using Eq.~(\ref{estimator}) with $\tilde{C}_{\ell}^{\textrm{exp}}(\beta)$ and the same equation but with $\tilde{C}_{\ell}^{\textrm{exp}}(0)$ will enable us to determine, in one full sky and in 20 cut sky simulations, whether the use of the estimator (\ref{estimator}), that is, neglecting the peculiar motion of the observer, is indeed a completely solid assumption or not.  
\begin{figure}[t]
\begin{center}
\begin{minipage}{60mm}
\includegraphics[width=65mm,height=50mm]{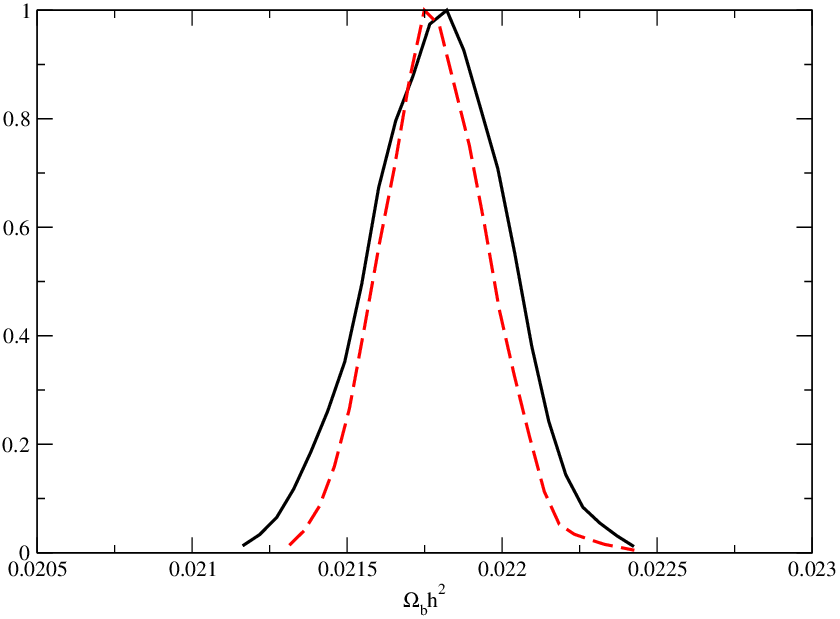}
\vspace{10mm}
\end{minipage}
\hspace{20 mm}
\begin{minipage}{60mm}
\includegraphics[width=65mm,height=50mm]{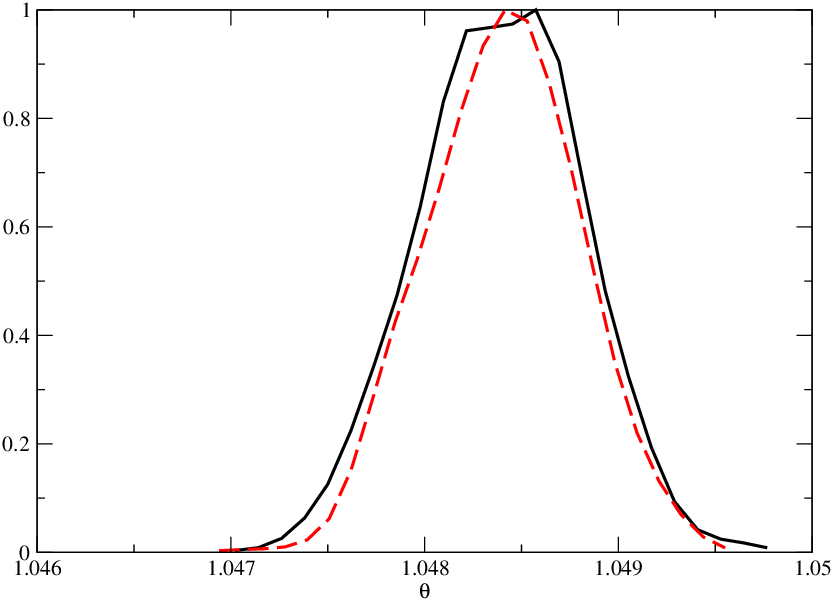}
\vspace{10mm}
\end{minipage}
\begin{minipage}{60mm}
\includegraphics[width=65mm,height=50mm]{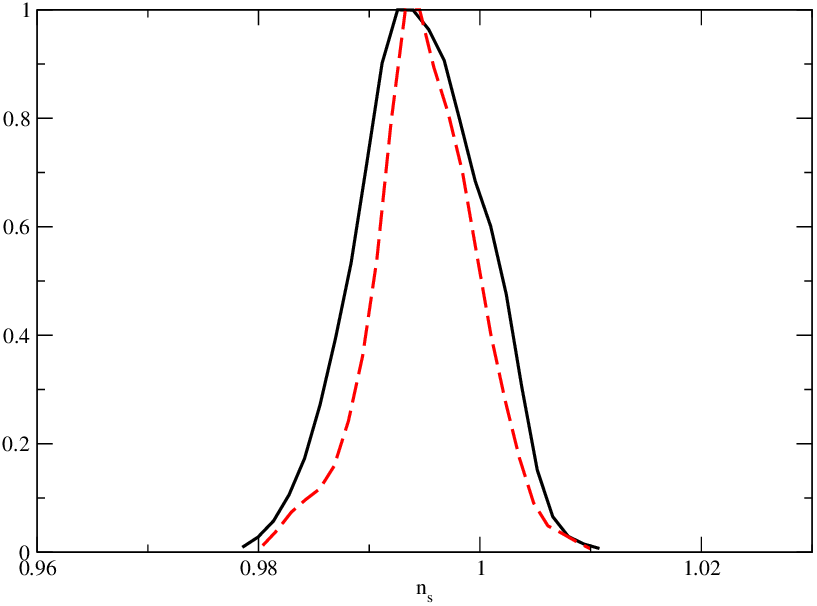}
\end{minipage}
\hspace{20mm}
\begin{minipage}{60mm}
\includegraphics[width=65mm,height=50mm]{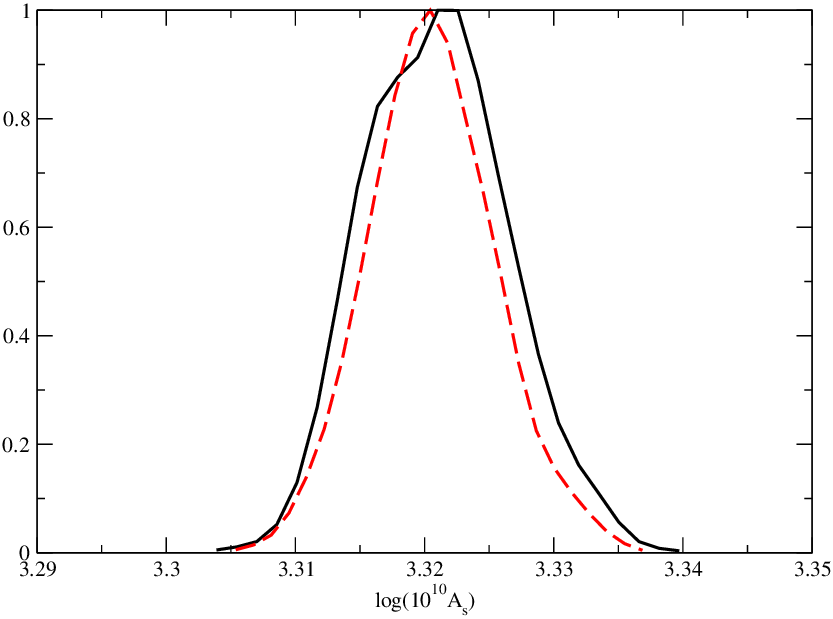}
\end{minipage}
\end{center}
\caption{Posterior probability density functions obtained from the simulation $\mathcal{S}$ using the associated {\it full sky} power spectra. The black curves refers to the analysis of a temperature map simulated in the CMB rest frame, while the red curves are associated with a run of CosmoMC performed using a temperature map simulated including aberration and Doppler effects. The experimental noise used in these figures is the one estimated for a Planck-like experiment.}
\label{S0}
\end{figure}

\subsection{Likelihood}
\label{like}
We will tackle the problem of reconstructing the parameters of our fiducial cosmological model from the data simulated in section \ref{mock} within a Bayesian framework. Bayesian methods combined with MCMC scanning techniques allow an effective comparison between observations and theoretical predictions. The outputs of this type of analyses are marginal posterior probability density functions (pdf) and credible intervals for the underlying model parameters (for a review on these subjects, see for instance Ref.~\cite{Trotta:2008qt}). In a Bayesian analysis the experimental input are encoded in the Likelihood function, which in our case will be a function of the model parameters and of the simulated data. We use here the Likelihood approximation introduced in Ref.~\cite{Hamimeche:2008ai}, according to which the Likelihood $\mathcal{L}$ can be written as follows 
\beq
-2\log\mathcal{L} \simeq \sum_{\ell\ell'} g(C^{\dagger}_{\ell}/C_{\ell}) \,C_{\ell}^{(f)} \,\left[M^{(f)}\right]^{-1}_{\ell\ell'}  \,C_{\ell'}^{(f)}\, g(C^{\dagger}_{\ell'}/C_{\ell'})
\eeq
where the function $g$ is explicitly given by
\label{likeap}
\beq
g(x) = \operatorname{sgn}(x-1)\sqrt{2(x-\ln(x)-1)} \,,
\eeq
 $\left[M^{(f)}\right]_{\ell\ell'}$ is the covariance matrix of the estimator (\ref{estimator}) evaluated at the fiducial model, $C_{\ell}^{(f)}$ is the angular power spectrum of the fiducial model and $C_{\ell}$ is the predicted angular power spectrum ({\it e.g.} computed with CAMB). For the fiducial model considered in this analysis, this matrix has been already estimated in Ref.~\cite{Hamimeche:2008ai} and can be downloaded in \cite{lewis}. In the presence of isotropic noise with known angular power spectrum $N_{\ell}$, it is straightforward to include its contribution to Eq.~(\ref{likeap}). It is in fact enough to sum to all the power spectra in Eq.~(\ref{likeap}) a noise contribution $N_{\ell}$. Concerning $N_{\ell}$, we implement in our analysis the same Planck noise simulated in Ref.~\cite{Hamimeche:2008ai}. 

Given a signal estimator $C^{\dagger}_{\ell}$ (see section \ref{mock}), a covariance matrix for it $\left[M^{(f)}\right]_{\ell\ell'}$ and a noise power spectrum $N_{\ell}$, the model parameter reconstruction can be performed with the CosmoMC program \cite{Lewis:2002ah}, using  these quantities. We restrict our analysis to multipoles $\ell>30$ to avoid complication related to the choice of the Likelihoods at small $\ell$. Finally, the window function used in this work and applied in the last call to the anafast program in the data simulation process (see section \ref{exp}) consists in an azimuthal cut of 20 degrees around the Galactic plane. This corresponds to a value of the $f_{sky}$ parameter equal to $0.826$. 
\begin{figure}[t]
\begin{center}
\begin{minipage}{60mm}
\includegraphics[width=65mm,height=50mm]{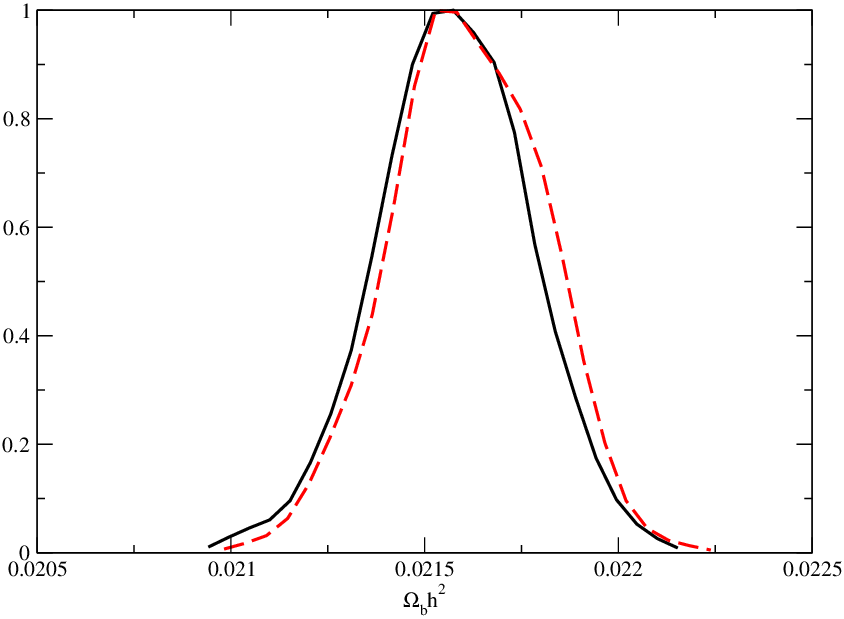}
\vspace{10mm}
\end{minipage}
\hspace{20mm}
\begin{minipage}{60mm}
\includegraphics[width=65mm,height=50mm]{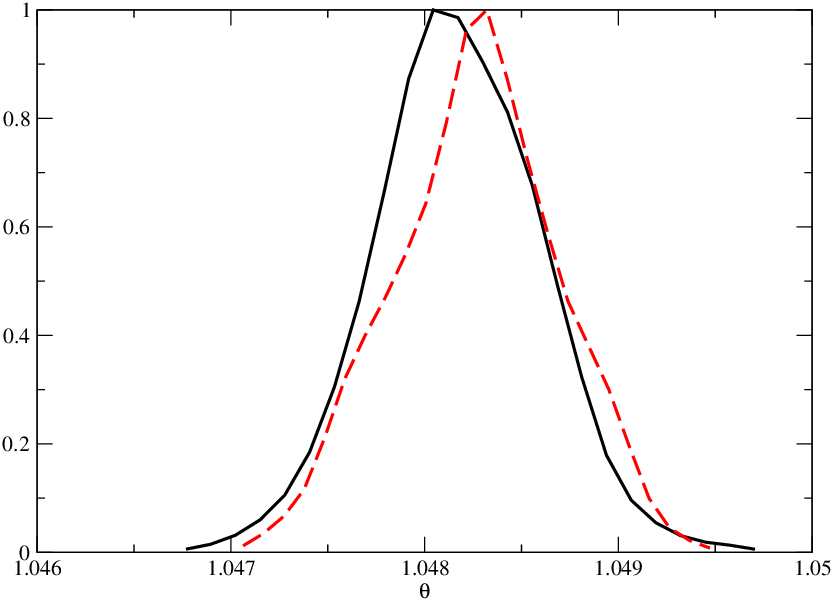}
\vspace{10mm}
\end{minipage}
\begin{minipage}{60mm}
\includegraphics[width=65mm,height=50mm]{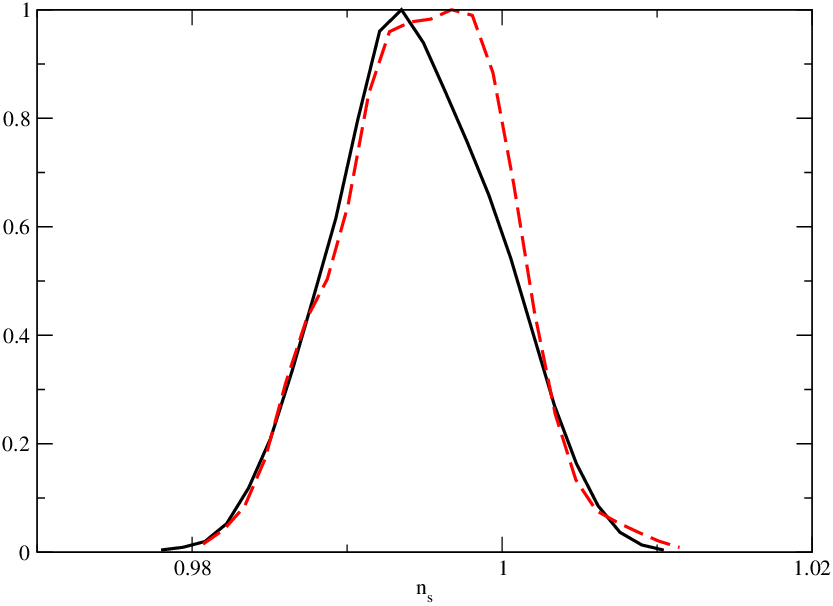}
\end{minipage}
\hspace{20mm}
\begin{minipage}{60mm}
\includegraphics[width=65mm,height=50mm]{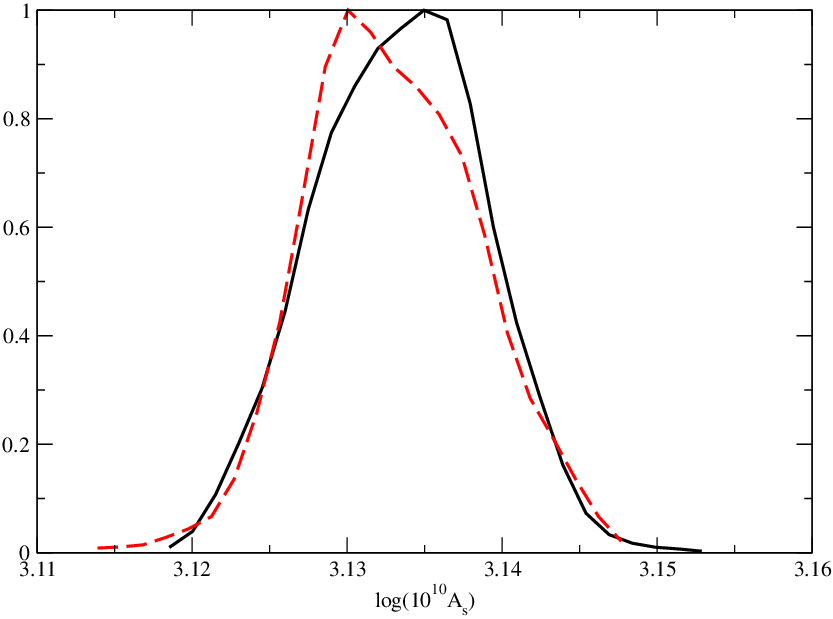}
\end{minipage}
\end{center}
\caption{Posterior probability density functions obtained from the simulation $\mathcal{S}$ using the associated {\it cut sky} power spectra. The black curves refers to the analysis of a temperature map simulated in the CMB rest frame, while the red curves are associated with a run of CosmoMC performed using a temperature map simulated including aberration and Doppler effects. The experimental noise used in these figures is the one estimated for a Planck-like experiment.}
\label{S1}
\end{figure}
\subsection{Full sky results}
We now compare the results of two cosmological parameter estimations performed using the two full sky power spectra associated with the simulation $\mathcal{S}$: one evaluated in the CMB rest frame and the other one obtained including aberration and Doppler. 
In the four panels of Fig.~\ref{S0} we plot the marginal posterior pdf's referring to four parameters of our toy cosmological model, namely $\Omega_{b} h^2$, $\theta$, $n_s$ and $A_s$. For every parameter we show the results obtained for the simulation $\mathcal{S}$ using $\tilde{C}_{\ell}^{\textrm{exp}}(0)$ (black lines) and $\tilde{C}_{\ell}^{\textrm{exp}}(\beta) $(red lines), in order to estimate the impact of aberration and Doppler on this parameter reconstruction.  More specifically, in every panel, the black solid line corresponds to the marginal posterior pdf derived from $\tilde{C}_{\ell}^{\textrm{exp}}(0)$. The red dashed line, instead, corresponds to the posterior pdf obtained from the dataset $\tilde{C}_{\ell}^{\textrm{exp}}(\beta)$. To quantify the impact of aberration and Doppler on cosmological parameters it is convenient to introduce the bias $\Delta \equiv \frac{(\Delta p)_{ab}}{(\Delta p)_{Planck}}$, where  $(\Delta p)_{ab}\equiv\langle p \rangle-\langle p' \rangle$, being $\langle p \rangle$ and $ \langle p' \rangle$ the mean values in the two frames (at rest and in the moving frame) for a parameter $p$, and $(\Delta p)_{Planck}$ is the error on that parameter quoted by the Planck collaboration~\cite{Ade:2013zuv}. We find that $\Delta$ is typically a few times $10^{-2}$.

\begin{figure}[t]
\begin{center}
\includegraphics[width=110mm,height=80mm]{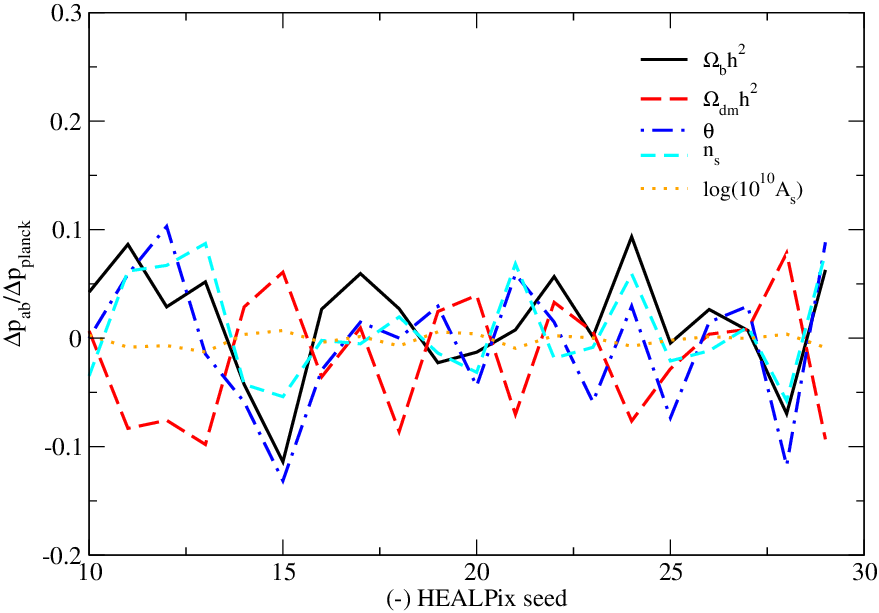}
\end{center}
\caption{$(\Delta p)_{ab}/(\Delta p)_{Planck}$ as a function of the Healpix seed used to generate the maps from which $(\Delta p)_{ab}$ has been derived. Different curves correspond to distinct cosmological parameters. In a few cases $(\Delta p)_{ab}$ is larger than the 10\% of  $(\Delta p)_{Planck}$, {\it i.e.}~the error quoted by the Planck collaboration for the corresponding parameter.}
\label{allparam}
\end{figure}
\subsection{Cut sky results}
We now present the outcome of our 20 cut sky simulations. Each simulation produced a pair of temperature maps: one in the CMB rest frame and one including Doppler and aberration. We analyzed these maps as explained above by means of the program CosmoMC. As in the case of a full sky analysis, in a single figure we report here, for one of these CosmoMC run, the relevant statistical outputs for the parameters $\Omega_{b} h^2$, $\theta$, $n_s$ and $A_s$. More precisely, Fig.~\ref{S1} shows the results obtained using the cut sky power spectra $\tilde{C}_{\ell}^{\textrm{exp}}(\beta)$ and $\tilde{C}_{\ell}^{\textrm{exp}}(0)$ associated with the simulation $\mathcal{S}$. We are using therefore the same HEALPix random seed previously used in the full sky analysis. Comparing the marginal posterior pdf's in this figure, we find that the statistical outputs of this analysis are mildly biased. This is the consequence of having analyzed the dataset $\tilde{C}_{\ell}^{\textrm{exp}}(\beta)$ assuming that the observer is comoving with the CMB, {\it i.e.} assuming therefore $\beta=0$. For this specific simulation, we also show in Table~\ref{Planck} for each parameter the means of the pdf's found using $\tilde{C}_{\ell}^{\textrm{exp}}(\beta)$ and $\tilde{C}_{\ell}^{\textrm{exp}}(0)$. For every parameter we also report the corresponding value of $\Delta$, namely the difference between the two means in units of $(\Delta p)_{Planck}$, {\it i.e.}~the error quoted by the Planck collaboration for that parameter~\cite{Ade:2013zuv}. For this specific Planck-like simulated experiment the means of the cosmological parameters could differ up to 10\% of the corresponding $(\Delta p)_{Planck}$. In particular the most affected parameters in this simulation are $\Omega_b h^2$ and $\theta$. This effect is rather small and, for a given pair $\tilde{C}_{\ell}^{\textrm{exp}}(\beta)$, $\tilde{C}_{\ell}^{\textrm{exp}}(0)$ associated with a certain {\it HEALPix random seed}, it might be simply due to the specific choice of the {\it CosmoMC random seed} made in the cosmological parameter estimation (different CosmoMC seeds have been used analyzing $\tilde{C}_{\ell}^{\textrm{exp}}(\beta)$ and $\tilde{C}_{\ell}^{\textrm{exp}}(0)$). We will come back to this aspect at the end of this section.

We then evaluated the previously defined bias $\Delta$ for the other 19 simulations at disposal and for every model parameter. The results of this study are shown in Fig.~\ref{allparam}. We find that varying the HEALPix seed (the simulation $\mathcal{S}$ corresponds to the HEALPix seed -15) the bias oscillates around zero. The standard deviation for $\Delta$ depends from the specific parameter considered, being equal to 0.05, 0.05, 0.06, 0.05, and 0.006 for respectively $\Omega_b h^2$, $\Omega_{dm} h^2$, $\theta$, $n_s$ and $\log(10^{10} A_s)$. Although a more careful analysis (eventually including polarization and considering a more realistic cosmological model) is necessary in order to achieve a more robust determination of the impact of aberration and Doppler on the cosmological parameter estimation, from the simulations discussed in this paper, we can already draw the conclusion that the common lore of neglecting the transformation law (\ref{boost}) in a CMB based parameter analysis is a safe assumption for a Planck-like sensitivity and sky-coverage, with a symmetric mask. 

To further corroborate this conclusion we focused on the HEALPix simulation $\mathcal{S}$ and analyzed the associated full-sky and cut-sky maps increasing the precision of the CosmoMC runs ({\it i.e.} lowering the value of the parameter MPI\_Limit\_Converge\_Err). Indeed the spectra $\tilde{C}_{\ell}^{\textrm{exp}}(\beta)$ and $\tilde{C}_{\ell}^{\textrm{exp}}(0)$ were analyzed with different random seeds of the CosmoMC program and it might be therefore that the 10\% bias identified in our previous analyses is at least partially related to this fact. By lowering the parameter MPI\_Limit\_Converge\_Err we can refine our previous estimations and mitigate spurious effects related to having analyzed $\tilde{C}_{\ell}^{\textrm{exp}}(\beta)$ and $\tilde{C}_{\ell}^{\textrm{exp}}(0)$ with different CosmoMC seeds. To this aim we computed the parameter bias $\Delta$ from $2 \times 8$ independent CosmoMC runs (these where performed with a lower value, compared to the runs shown in Fig.~\ref{allparam}, for the parameter MPI\_Limit\_Converge\_Err); 8 runs associated with $\tilde{C}_{\ell}^{\textrm{exp}}(\beta)$ and 8 with $\tilde{C}_{\ell}^{\textrm{exp}}(0)$. We found in this way eight different determinations of $\Delta$ for each model parameter. By computing mean and standard deviation of the parameters we actually find no significant difference between the simulations with $\beta=0$ and with $\beta\neq 0$. In all cases $\Delta$ is lower than a few percent, a result which allows to safely neglect the transformation law (\ref{boost}) in a CMB based parameter analysis performed with a Planck-like sensitivity and sky-coverage. Finally we also checked whether this conclusion is modified when asymmetric cut-sky are considered finding however that also in this case aberration and Doppler effects induce only a mild bias in the cosmological parameter estimation.

\begin{table}
    \centering
    \begin{tabular}{lccccc}
    \toprule
    Parameter         & Mean ($\beta=0$) & $\sigma$ ($\beta=0$) &  Mean ($\beta\neq0$) & $ \sigma$ ($\beta\neq0$) & $|\Delta|$ (Planck) \\
    \midrule                                          
    $\Omega_{b} h^2$   & 0.02157 & 1.89$\times10^{-4}$ & 0.02161 & 1.88$\times10^{-4}$ & \bf{0.11} \\
    $\Omega_{dm} h^2$   & 0.1214 & 2.12$\times10^{-3}$ & 0.1212 & 2.14$\times10^{-3}$ & \bf{0.06} \\
    $\theta$ & 1.04817 & 4.12$\times10^{-4}$ & 104826 & 4.14$\times10^{-4}$ & \bf{0.13} \\
    $n_s$  &  0.9944 &  5.04$\times10^{-3}$ & 0.9950 &  4.99$\times10^{-3}$ &  \bf{0.06} \\
    $\log(10^{10} A_{s})$  & 3.1334 & 5.22$\times10^{-3}$ & 3.1329 & 5.32$\times10^{-3}$ &  \bf{0.007} \\
    \bottomrule
    \end{tabular}
    \caption{\label{Planck} Means and standard deviations for the cosmological model parameters derived by analyzing the datasets $\tilde{C}_{\ell}^{\textrm{exp}}(0)$ and $\tilde{C}_{\ell}^{\textrm{exp}}(\beta)$ of the simulation $\mathcal{S}$ using a Planck-like noise. For every parameter we also list in the last column the absolute value of $\Delta$, namely the difference of the two means (second and fourth column) divided by $(\Delta p)_{Planck}$, {\it i.e.}~the error associated with the corresponding parameter quoted by the Planck collaboration~\cite{Ade:2013zuv}. $\Delta$ provides a quantitative estimation of the error associated with neglecting aberration and Doppler effects.}
\end{table}

\section{Analytical estimates}
\label{fitting}

In this section we present how to perform analytical checks and estimates of the results that we have found numerically.  In fact in~\cite{Notari:2011sb}  fitting formulas given in terms of Bessel functions have been found to reproduce to high accuracy ($0.2\%$) the exact coefficients of eq.~(\ref{abkernel}) at least up to $\ell=700$.
In particular for the Temperature coefficients they read:
\begin{equation}\label{eq:non-linear-fit-general}
\begin{aligned}
    K_{\ell-n\, \ell\, m}^X   &\;\simeq\; J_n\!\left(\!-2\, \beta \left[\prod_{k=0}^{n-1} \big[(\ell-k) \;G_{\ell-k\, m} \big]\right]^{1/n} \right), \\
    K_{\ell+n\, \ell\, m}^X   &\;\simeq\; J_n\!\left(\,2\, \beta \left[\prod_{k=1}^n \big[(\ell+k) \;G_{\ell+k\, m} \big]\right]^{1/n} \right),
\end{aligned}
\end{equation}
where $J_n$ is the Bessel function of the $n$-th kind and where
\begin{align}\label{eq:challinor-fun}
   G_{\ell\, m} \;\equiv\; \sqrt{\frac{\ell^2-m^2}{4\ell^2-1} }\, ,
\end{align}
while for $n=0$: 
\begin{equation}\label{eq:non-linear-fit-n0}
\begin{aligned}
    K_{\ell\, \ell\, m}^X  \;\simeq\; J_0\!\Bigg(&  \beta\sqrt{2} \,\bigg[ \!-(\ell+1)\,(\ell+2)\,({}_sG_{\ell+1\, m})^2\,-\ell\,(\ell-1) \,({}_sG_{\ell\, m})^2 + \ell(\ell+1) - m^2 +1 \bigg]^\frac{1}{2} \Bigg).
\end{aligned}
\end{equation}

We have checked that we can reproduce to high accuracy the same $C_{\ell}$'s of our numerical program, using such fitting formulas to transform a set of randomly generated $a_{\ell m}$'s, as long as we include mixing with a sufficiently large number of neighbours in eq.~(\ref{matrixalm}). For instance, including 2 neighbours we obtain agreement between the two methods of the order $(C_{\ell}^{\rm exp}-C_{\ell}^{\rm fit})/C_{\ell}^{\rm exp}\approx 10^{-6}$ at $\ell=100$, where $C_{\ell}^{\rm exp}$ are the numerical results and $C_{\ell}^{\rm fit}$ are the ones obtained with fitting functions. The precision goes down to $10^{-7}$  when including  3 or more neighbours. 
At higher $\ell$ (between 1000 and 2000) the error rapidly grows with $\ell$ if we consider only two neighbours. When considering 4 neighbours the error still grows with $\ell$ but it is always smaller than $5\times 10^{-4}$, upto $\ell=2000$, while with $5$ neighbours it goes down to $ 10^{-4}$.

The fact that two completely independent methods, namely our numerical procedure and the Bessel fitting functions, give results which coincide with high precision provides a check  in favor of the correctness of our numerical approach.

\section{Conclusions}
\label{conclusion}

In this paper we have considered the effect of our peculiar velocity on  CMB maps with 
Planck-like resolution and sky-coverage in particular focusing on the power spectrum and on cosmological parameter estimation. We have assumed a peculiar velocity of $\beta=1.23 \times 10^{-3} c$ in the direction of the CMB dipole and we have applied a boost transformation (Eqs.(\ref{aberration}) and (\ref{boost})) to this frame directly in pixel space on simulated maps, rather than on the $a_{\ell m}$'s, which allows us to overcome the challenging problem of evaluating efficiently a large number of integrals of highly oscillating Legendre polynomials.
For this purpose we have simulated 20 temperature maps with $\ell < 2000$ and with a noise which reproduces the Planck satellite resolution as a random  realization of a fiducial five-parameters cosmological model, for which a simple Likelihood functions has been tested in~\cite{Hamimeche:2008ai}. For simplicity we have also discarded the first 30 multipoles, which are usually treated with a different likelihood. Finally we have performed, both for full sky and for a cut-sky configurations, an analysis of the maps by running a MCMC on each map with or without the aberration plus Doppler effect. For cut-sky we have used a cut of 20 degrees around the Galactic
plane.

%We have also checked that using fitting functions for the highly oscillating integrals as in~\cite{Notari:2011sb} we can reconstruct the aberrated $C_{\ell}$'s analytically with high precision and we have shown that such precision is enough to lead to the same results for the bias on Cosmological parameters.

As a conclusion we find that neglecting aberration and Doppler would induce a negligible bias on a CMB based cosmological parameter estimation for Planck-like resolution and sky-coverage.
In fact we find that the mean values for all the cosmological parameters shift by less than the 10\% of the error quoted by the Planck collaboration for that parameter, which is within the intrinsic uncertainty of the MCMC parameter estimation. While our analysis is not exhaustive because of many simplifications, we have shown that it is harmless to ignore aberration and Doppler in order to do cosmological parameter estimations for high resolution experiments such as Planck.  
%and even for WMAP, and that therefore it would be appropriate before analyzing the experimental maps to apply a Lorentz transformation which transforms  them back to the CMB rest frame. In order to do this it is crucial however to rely on the assumption that our velocity is approximately given by the CMB dipole, unless it can be measured to some extent in other ways, as proposed in~\cite{Amendola:2010ty,Notari:2011sb}.

\acknowledgments
Some of the results in this paper have been derived using the HEALPix package \cite{Gorski:2004by}. We also acknowledge the use of the CosmoMC code \cite{Lewis:2002ah}. We thank Miguel Quartin, Antony Lewis, Paolo Natoli and Gianluca Polenta for useful discussions and comments.

\end{document}